# Strongly Enhanced Current Densities in Superconducting Coated Conductors of YBa$_2$Cu$_3$O$_{7-x}$ + BaZrO$_3$

J. L. MacManus-Driscoll[1,2], S. R. Foltyn[1], Q. X. Jia[1], H. Wang[1], A. Serquis[1], L. Civale[1], B. Maiorov[1], M.E. Hawley[1], M.P. Maley[1] and D. E. Peterson[1]

[1] Superconductivity Technology Center, Los Alamos National Laboratory, Los Alamos, NM 87545, U.S.A.

[2] on leave from Dept. of Materials Science and Metallurgy, University of Cambridge, Pembroke St., Cambridge, CB2 3QZ, U.K.

There are numerous potential applications for superconducting tapes, based on YBa$_2$Cu$_3$O$_{7-x}$ (YBCO) films coated onto metallic substrates [1]. A long established goal of more than 15 years has been to understand the magnetic flux pinning mechanisms which allow films to maintain high current densities out to high magnetic fields [2]. In fact, films carry 1-2 orders of magnitude higher current densities than any other form of the material [3]. For this reason, the idea of further improving pinning has received little attention. Now that commercialisation of conductors is much closer, for both better performance and lower fabrication costs, an important goal is to achieve enhanced pinning in a practical way. In this work, we demonstrate a simple and industrially scaleable route which yields a 1.5 to 5-fold improvement in the in-field current densities of already-high-quality conductors.

The sources of enhanced pinning in vapour-grown YBCO films are the natural point, line and volume imperfections, probably the most significant of these being the dislocations perpendicular to the substrate plane of film [4]. In terms of dimensionality, dislocations are nearly ideal for pinning magnetic flux lines. However, the density of dislocations is dominated by the growth island size and their spacing is estimated to be rather large (~100 nm-~500 nm) [3, 5]. To increase dislocation density, an obvious way would be to decrease the island size which the dislocations bound [6], e.g. by reducing growth temperature, but this is non-trivial because the crystalline quality of the film would be compromised.

Heavy ion irradiation has been shown to reduce vortex mobility [7, 8] but it is impractical for treatment of coated conductors. Other work involving growth of films on mis-cut single crystal substrates has demonstrated that introduction of columnar growth defects improves $J_c$ (77K) by up to 50%, but only at a particular field orientation and magnitude [9]. Other ideas for improving pinning are introduction of defects by multi-layering or addition of particles on the substrate surface [5,10]. Using these methods some improvements appear possible in thin films on single crystal substrates.

In this work, prompted by our earlier report that suggested the possibility of enhanced pinning in the presence of epitaxial second phases [11], we study BaZrO$_3$ additions to YBCO. The main reasons for the choice of BaZrO$_3$ are a) while it can grow heteroepitaxially with YBCO it has a large lattice mismatch (~ 9%) so strain between the phases could introduce defects for enhanced pinning, b) it is a high melting temperature phase and so growth kinetics should be slow, leading to small particles, and c) Zr does not substitute in the YBCO structure [12]. Indeed, single crystals of YBCO are often grown in $BaZrO_3$-coated crucibles [13]. $BaZrO_3$ has also been previously investigated as a pinning centre in bulk, melt processed YBCO [14, 15]. However, it was found that the $BaZrO_3$ agglomerated at the growth fronts of the grains, and, because of heteroepitaxial matching with the YBCO lattice, it acted as a seed to nucleate multiple grains instead of a single domain.

We show that nano-particles of $BaZrO_3$ grow heteroepitaxially within laser ablated YBCO films. These particles are easily incorporated in the films from the source target of a ceramic $BaZrO_3$/YBCO mixture. The particles lead to significant improvement in the in-field $J_c$ (at 75.5K) in films both on single crystal substrates and on practical, buffered metallic substrates.

Ceramic targets were prepared from a) pure YBCO, and b) YBCO+5mol.% $BaZrO_3$. Commercial YBCO powder was used, as well as 99.99% pure powders of $Ba(NO)_3$ and $ZrO_2$. The powders were mixed, ground, pressed and then sintered at 950°C in flowing oxygen gas. The targets were ablated using pulsed laser deposition (PLD) with a KrF excimer laser ($\lambda$= 248nm), at a repetition rate of 10 Hz. All of the depositions were carried out at the same substrate-to-target distance of 5 cm and an oxygen pressure of 200 mTorr. The substrates used were either single crystal $SrTiO_3$ (STO), $SrTiO_3$-buffered MgO single crystals, or $SrTiO_3$-buffered-ion beam assisted deposition-MgO-on-Hastelloy [16], hitherto referred to as IBAD-MgO. After deposition at 760-790°C, samples were cooled to room temperature in $O_2$ at 300 Torr. Ten different YBCO+$BaZrO_3$ samples were grown with thicknesses in the range 0.5-1.7μm.

For all the samples, inductive critical temperature ($T_c$) and transport critical current density, $J_c$, measurements in self field at liquid $N_2$ temperatures were made. Further transport measurements were conducted on some of the samples in liquid $N_2$, in a magnetic field rotated in a plane perpendicular to the plane of the film but always normal to the current (maximum Lorentz force configuration). Microstructural characterization was carried out by x-ray diffractometry (XRD), atomic force microscopy (AFM), and cross-sectional transmission electron microscopy (TEM).

Table 1 shows the measured data for pure YBCO films and several YBCO+$BaZrO_3$ films. The $T_c$'s and self-field $J_c$'s of some of the YBCO+$BaZrO_3$ samples are slightly lower than for the pure YBCO. Figure 1 a shows $J_c/J_c^{sf}$ ( where $J_c^{sf}$ is self-field $J_c$) versus magnetic field (H || c) for a YBCO+$BaZrO_3$ film compared to two pure YBCO films of different thickness on STO single crystals. Figure 1 b shows $J_c$ versus field (H || c) for a YBCO+$BaZrO_3$ film compared to a pure YBCO film, both on IBAD-MgO. Figure 1 a shows that pure YBCO films of different thickness give very similar forms of the $J_c/J_c^{sf}$ versus field (H || c) curve. We always find this to be the case for samples of different thickness (in the range 0.5-2 μm) if the $T_c$ is the same, and if samples are compared on single crystal or on the s*ame* batch of IBAD-MgO (the *same* batch of IBAD-MgO was used for the samples in Table 1).

The striking result of Figure 1 is the upwards shift in both normalised $J_c$ and $J_c$ for the YBCO+BaZrO$_3$ samples. Identical behaviours were observed for the two other samples measured on single crystal substrates (13 and 83). Even though the YBCO+BaZrO$_3$ samples of Figure 1 have slightly lower $T_c$'s than the pure ones, the in-field $J_c$'s are improved significantly. Hence, the BaZrO$_3$ addition has clearly increased the irreversibility field. On both the single crystal and IBAD-MgO substrates, accounting for the thickness differences between samples the $J_c$ values are around a factor of 1.5-2 higher over a wide field range (~1-5T) and they increase to a factor of around 5 higher at 7T. In ~1μm thick YBCO+BaZrO$_3$ films on IBAD, $J_c$'s remain in excess of 0.1MA.cm$^{-2}$ at 4.5T.

The inset of Figure 1 a shows the angular dependence of $J_c$ measured in a field of 1T for the samples on STO single crystal. A shift upwards in the relative height of the c-axis angular peak is observed for YBCO+BaZrO$_3$ compared to the pure YBCO, indicative of strong pinning defects along the c-axis in the YBCO+BaZrO$_3$ film [17]. In fact, the $J_c$ is increased substantially compared to the pure sample across the angular range from 0° to ~80°. The same trend is also observed for samples on IBAD-MgO. The result is important for applications of coated conductors, since the magnetic field will rarely be constrained to a single orientation. Previous measurements of the $J_c$ (H ∥ c) dependences for pure YBCO films on for the samples on STO single crystal STO and IBAD-MgO have been shown to be very similar, suggesting a similar pinning mechanism on different substrates.

The inset of Figure 1 b shows the pinning force ($F_p = \mu_o H J_c$) normalised by the maximum pinning force measured for sample 35 ($F_p^{max}(\#35)$), versus field for the samples on STO single crystal. Despite the lower $T_c$ of the YBCO+BaZrO$_3$ film (35), the pinning force is significantly higher compared to the pure YBCO films (26 and 60).

X-ray diffractograms of YBCO+BaZrO$_3$ films (not shown) show peaks belonging to (00l) YBCO plus an additional, broad, low intensity peak centred around 2θ = 42.7° which is consistent with (200)BaZrO$_3$, and/or (200)Ba$_2$Zr$_{2-x}$Y$_x$O$_6$, and/or (400) Ba$_2$ZrYO$_6$ [18]. Surprisingly, the Ba$_2$ZrYO$_6$ phase has not been widely reported previously, i.e. the assumption has generally been that BaZrO$_3$ does not react with YBCO. However, the breadth and position of the second phase peak suggests that Ba$_2$Zr$_{2-x}$Y$_x$O$_6$ particles with a range of x values are present. As shown from x-ray phi scans (Figure 2) of the YBCO (103/110) and BaZrO$_3$ (110) peaks at χ=45° and 2θ=32.8° and 30.2°, respectively, the particles are in-plane aligned cube-on-cube with the YBCO.

Figure 3 shows atomic force micrographs for a YBCO+BaZrO$_3$ film on STO. Nano-particles are observed distributed across the film surface. These particles are not observed in pure YBCO films. The phase contrast image of b) indicates that the particles are composed of a phase different from YBCO. Cross sectional low magnification and high resolution TEM micrographs (Figures 4a and b, respectively) confirm the presence of nano-particles embedded in the YBCO lattice. In addition, c-aligned, misfit edge dislocations of spacing <50 nm are present in the YBCO. Some of these dislocations are indicated by black arrows. The minimum density of columnar defects is at least 400 $\mu m^{-2}$, compared to ~80 $\mu m^{-2}$ previously observed in YBCO films grown by PLD [5]. The dislocations form a family of correlated linear defects that should produce substantial uniaxial pinning along the c direction, consistent with the observed enhancement of the c axis peak in $J_c$ (inset of Figure 1). It appears that the dislocations form as a result of the lattice misfit between the particles and matrix. Based on the film and particle lattice parameters of ~3.85 Å and 4.23Å, respectively, the level of strain is around 9.4% [19]. The high defect density is consistent with the reduced $T_c$'s and self-field $J_c$'s of some of the samples. While the 'c'-axis dislocations are the source of the enhanced directional pinning, it is possible that some random pinning may also arise directly from the $BaZrO_3$ particles.

Lattice images (Figure 4c) and Fast Fourier Transformations of the particles indicated a cubic structure, with lattice parameter a ≈ 4.23Å, consistent with $BaZr_{1-x}Y_xO_3$. Owing to the small sizes of the particles, it was not possible to conclusively determine their composition. A typical particle size distribution measured over a 1$\mu m^2$ area is shown in Figure 4d. The nano-particles ranged in size from 5 nm to 100 nm with a modal particle size of 10nm.

In summary, by implementing a straightforward and inexpensive target compositional modification, $J_c$ enhancements of up to a factor of 5 (depending on field) at liquid $N_2$ temperatures are achieved in a reproducible way on both single crystal and buffered metallic substrates. Since only one type of heteroepitaxial addition ($BaZrO_3$) and only one concentration (5 mol. %) were studied, it is likely that other concentrations and different heteroepitaxial second phase additions will lead to yet greater enhancements in pinning.

**Acknowledgments**

The work was supported by the Office of Energy Efficiency and Renewable Energy, U.S. Department of Energy. We thank Dr. John Durrell, Univ. of Cambridge. for helpful discussions.

**Competing Financial Interests**

The authors declare that they have no competing financial interests.

**Figure Captions**

**Figure 1: Crtitical current density at 75.5K versus magnetic field applied parallel to 'c' axis for pure YBCO and YBCO+BaZrO$_3$ films.** $J_c/J_c^{sf}$ is shown in a) for single crystal SrTiO$_3$ substrates and $J_c$ is shown in b) for IBAD-MgO substrates. Open points are for pure YBCO and closed points are for YBCO+BaZrO$_3$. Inset to a) shows angular dependence of $J_c$ at 1T with sample 60 data multiplied by 0.75, and inset to b) shows $F_p/F_p^{max}$(#35) versus magnetic field applied parallel to 'c' for the samples on single crystal STO.

**Figure 2: x-ray plot showing in-plane alignment of YBCO and BaZrO$_3$ particles.** phi scans at $\chi=45°$. $2\theta=30.2°$ corresponding to the BaZrO$_3$ (110) peak, and $2\theta=32.8°$ corresponding to the YBCO (103)/(110) peak.

**Figure 3: Micrographs of YBCO+BaZrO$_3$ films grown on single crystal SrTiO$_3$ showing surface nano-particles.** a) atomic force micrograph, and b) phase contrast micrograph.

**Figure 4: Transmission electron micrograph information showing nano-particles of BaZrO$_3$ in YBCO+BaZrO$_3$ films on SrTiO$_3$-buffered MgO single crystals.** Some of the nano-particles have been labelled using white arrows and some of the columnar defects by black arrows. a) high magnification micrograph, b) lower magnification micrograph, c) image of nano-particle showing lattice fringes consistent with a cubic lattice, d) histogram showing distribution of particle sizes.

| Sample number | $T_c$ (breadth), K | Self field $J_c$(75.5K) | Thickness (μm) | Substrate |
|---|---|---|---|---|
| Pure YBCO | | | | |
| 26 | 91.7(0.7) | 2.4 | 1.55 | STO |
| 60 | 91.5(0.5) | 2.6 | 1.0 | STO |
| 87 | 92(3) | 2.3 | 1.2 | STO on IBAD |
| YBCO+BaZrO$_3$ | | | | |
| 12 | 91 (1) | 2.4 | 0.5 | STO |
| 13 | 92 (2) | 2.3 | 1.7 | STO |
| 30 | 91.5 (1) | >2.2*† | 1.3 | STO |
| 35 | 89.5 (2.5) | 2.0 | 0.75 | STO |
| 37 | 87.8 (0.5) | 2.2 | 1.0 | STO |
| 83 | 90 (1) | >1.8* | 1.3 | STO on MgO |
| 91 | 89 (1) | 1.5 | 1.2 | STO on MgO |
| 94A | 87.5 (1.5) | 1.7 | 0.9 | STO on IBAD |
| 94B | 88 (1) | 2 | 1.0 | STO on MgO |
| 95B | 88.7 (2) | >2.6* | 1.2 | STO on MgO |

\* Current carried in bridge exceeded measurement limit of the current source of 10A
† Bridge blew during measurement

Table 1: Sample data for reference YBCO films, and for YBCO+BaZrO$_3$ films on single crystal SrTiO$_3$ (STO), MgO-buffered single crystal STO, and STO-buffered-IBAD-MgO

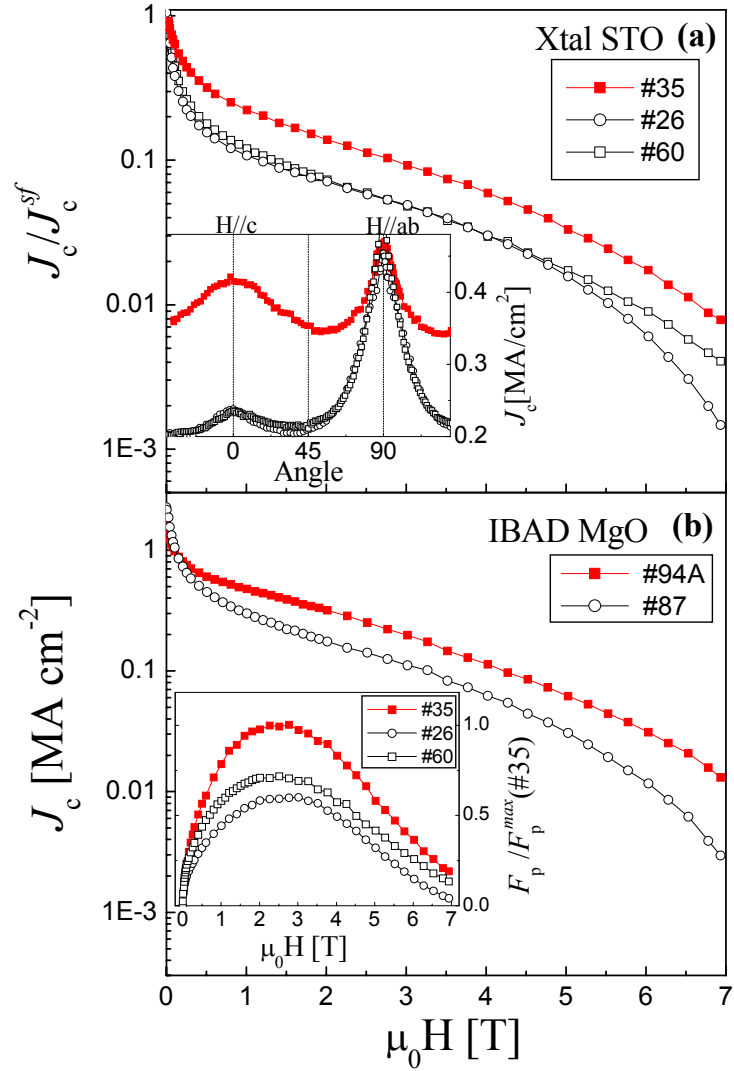

Figure 1 J. L. MacManus-Driscoll *et al*

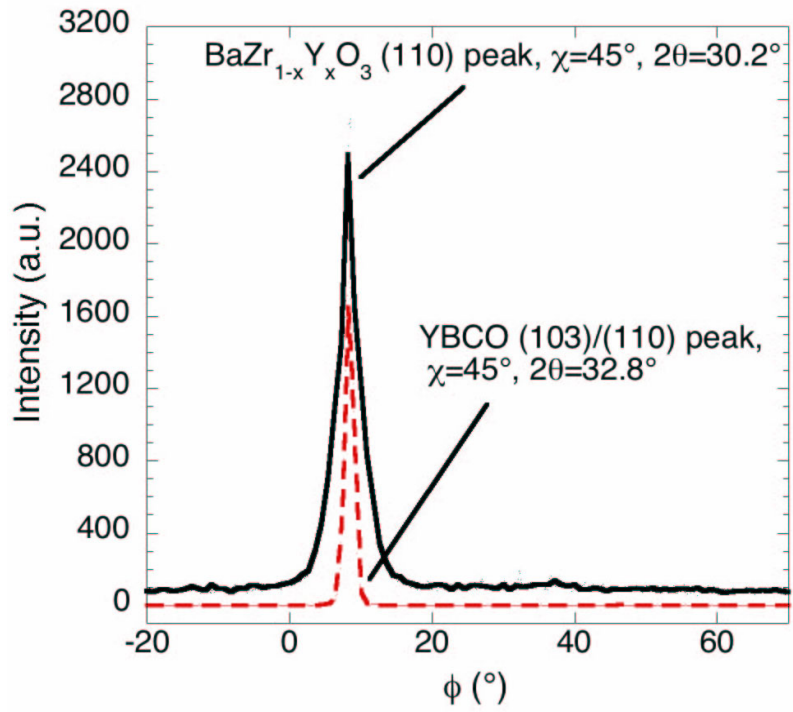

Figure 2 J. L. MacManus-Driscoll *et al*

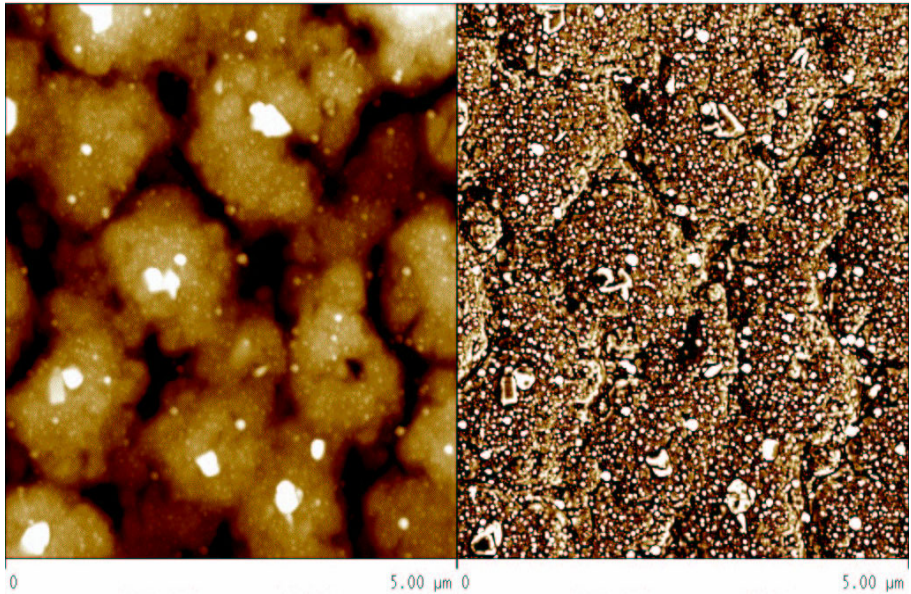

Figure 3 J. L. MacManus-Driscoll *et al*

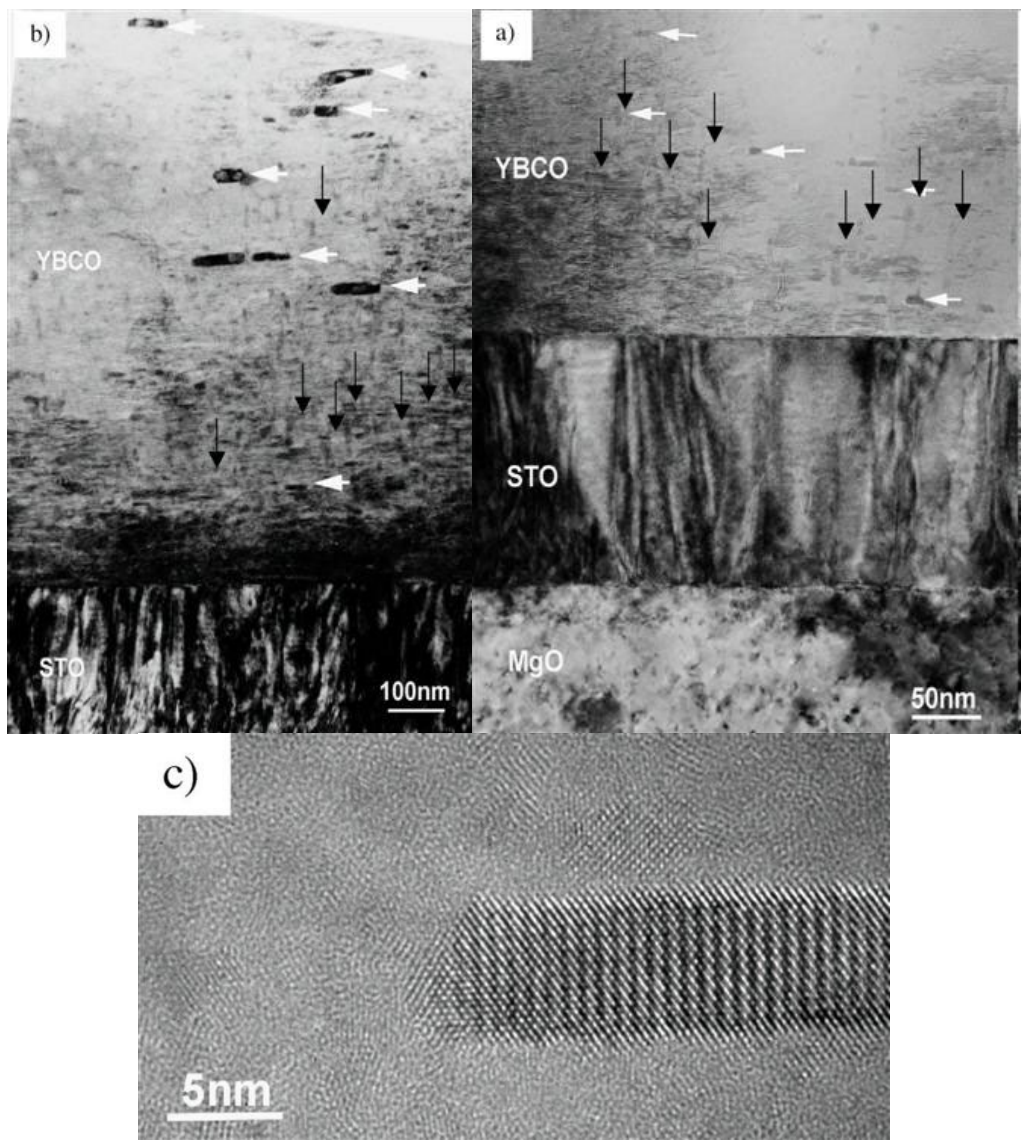

Figure 4 J. L. MacManus-Driscoll *et al*